\documentclass[12pt]{article}

\font\grande=cmr9.5 scaled \magstep4
\font\medio=cmr9.5 scaled \magstep2
\outer\def\beginsection#1\par{\medbreak\bigskip
      \message{#1}\leftline{\bf#1}\nobreak\medskip
\vskip-\parskip
      \noindent}
\usepackage{graphicx} 
\begin{document}
\bibliographystyle {unsrt}

\titlepage

\begin{flushright}
\end{flushright}

\vspace{10mm}
\begin{center}
{\grande Averaged Energy Conditions and Bouncing Universes}\\
\vspace{1.5cm}
 Massimo Giovannini
 \footnote{Electronic address: massimo.giovannini@cern.ch}\\
\vspace{1cm}
{{\sl Department of Physics, 
Theory Division, CERN, 1211 Geneva 23, Switzerland }}\\
\vspace{0.5cm}
{{\sl INFN, Section of Milan-Bicocca, 20126 Milan, Italy}}
\vspace*{0.5cm}
\end{center}

\vskip 0.5cm
\centerline{\medio  Abstract}
The dynamics of bouncing universes is characterized by violating certain
coordinate invariant restrictions on the total energy-momentum 
tensor, customarily referred to as energy conditions. Although there could be 
epochs where the null energy condition is locally violated, 
it may perhaps be enforced in an averaged sense. Explicit examples of this 
possibility are investigated in different frameworks. 
\vskip 0.5cm

\noindent

\vspace{5mm}

\vfill
\newpage
Since the first WMAP releases, the upper limits on the tensor to scalar ratio $r_{T}$ 
have been steadily decreasing so that it is reasonable to admit, with a fair degree of confidence, 
 that $r_{T}$ must not exceed one tenth at a typical pivot scale of $0.002\, \mathrm{Mpc}^{-1}$ 
 \cite{WMAP}. The conventional inflationary scenarios (broadly compatible with the latter upper limit) 
 would imply a set of Cauchy data characterized by a minute energy density of the inflaton in Planck units \cite{WMAP}.
 The current bounds cannot exclude even smaller values of  $r_{T} $ (e.g. of the order of $10^{-5}$). 
 Since bouncing universes naturally lead to extremely small tensor to scalar 
 ratios \cite{boun0} they might represent a plausible completion of the protoinflationary dynamics 
 (see, for instance, \cite{boun1} for an incomplete list of reviews involving, directly or indirectly, 
 bouncing scenarios).

Bouncing universes are sometimes ruled out (or in) by appealing to various coordinate-invariant restrictions on the total 
energy-momentum tensor $T_{\mu\nu}$. These restrictions are conventionally referred to as 
energy conditions \cite{boun1,boun2}. 
In various classes of theories, bouncing solutions violate the null energy 
condition\footnote{The contraction of arbitrary time-like or null vectors with $T_{\mu\nu}$ leads to a number of different scalar 
functions playing a relevant role in various singularity theorems \cite{boun2}. 
Besides the null energy condition, the weak energy condition stipulates that 
$T_{\mu\nu} q^{\mu} q^{\nu} \geq 0$ where $q^{\mu}$ is a 
time-like vector (i.e. $q^{\mu} q^{\nu} g_{\mu\nu} =1$ ).  Conversely, the strong energy condition 
implies that $T_{\mu\nu} p^{\mu} p^{\nu} \geq T^{\lambda}_{\lambda} p^{\sigma} p_{\sigma}/2$
where $p^{\mu}$ is now a non-spacelike vector (i.e. either timelike or null).} demanding that $T_{\mu\nu} k^{\mu} k^{\nu} \geq 0$ 
where $k^{\mu}$ is a null vector (i.e. $g_{\mu\nu} k^{\mu} k^{\nu} =0$ and $g_{\mu\nu}$ denotes the four-dimensional metric tensor). 
Whenever the null energy condition is strictly enforced, the occurrence of bouncing universes is prevented, 
at least in the scenarios formulated within the conventional 
general relativistic dynamics. Indeed, if this is the  case, the first (cosmic) time 
derivative of the Hubble rate is always negative semidefinite and cannot change sign. The same 
kind of restrictions appear when applying  effective field theory methods to the analysis of single field cosmological models
where the dependence of the action on the scalar field can be constrained through the energy conditions. 
The effective field theory approach (originally developed in the framework of single-field inflationary scenarios 
 \cite{boun3}) can be translated into the case of bouncing dynamics and, in this context, 
 the enforcement of the null energy condition has been used to rule out a large class of 
 realizations of effective bouncing scenarios \cite{boun4}.  The question we ought to address 
 in this paper is the following: even if the null energy condition is violated locally in a given bouncing Universe, 
 can it be satisfied in average?  While there is no general proof of this statement we shall show that the answer 
 can be  affirmative at least in few explicit examples.

Although there are regions where the null energy condition can be locally violated, it may perhaps be 
satisfied in some averaged sense: this is, in a nutshell, the conservative suggestion conveyed hereunder in connection with the bouncing 
dynamics. The notion of averaged energy conditions (originally introduced by Ford and subsequently sharpened 
in a series of papers \cite{boun5}) aims at preventing the violation of the second law of thermodynamics 
and plays also a relevant role in the derivation of the so-called quantum interest conjecture. 
So far averaged energy conditions have not been specifically analyzed in the context 
of bouncing scenarios: the goal of the present investigation is to fill this gap, at least partially. 
For the present goals, a direct way of introducing the averaged null energy condition is to consider 
the following integral \cite{boun5}
\begin{equation}
 \int_{\gamma} T_{\mu\nu} k^{\mu} k^{\nu} \, d\lambda \geq 0,
 \label{AC}
 \end{equation}
 where $k^{\mu} = d x^{\mu}/d\lambda$ is the tangent vector to the null geodesic and $\lambda$ 
 is an affine parameter with respect to which the tangent vector to the geodesic is defined; in Eq. (\ref{AC}) 
 $\gamma$ denotes a null geodesic. In general relativity the integral condition (\ref{AC}) can also be phrased by 
 substituting $T_{\mu\nu}$ with the Ricci tensor $R_{\mu\nu}$: 
 the two conditions are in fact equivalent (up to numerical factors)
by definition of null vector and thanks to the field equations\footnote{The averaged null energy condition measures the degree of violation 
of the null energy condition. It has been argued (see eg. g. second paper of Ref. \cite{boun5}) that various general relativistic 
results can be demonstrated without requiring that energy conditions 
are satisfied locally, but only in an averaged sense. }.
 
To assess the restrictions implied by Eq. (\ref{AC}), 
the case of conformally flat Friedmann-Robertson-Walker 
geometries shall be preferentially considered but the possible extensions of this study to open or closed 
universes seem very plausible; the metric tensor shall then be expressed, for the present ends,  
as\footnote{We are dealing here with the case where the spatial curvature is absent (this is an immediate consequence of conformal flatness). 
In the presence of spatial curvature the bouncing behaviour may arise 
without an explicit violation of the null energy condition, as in the case of the de Sitter bounce which may also arise 
when the total energy-momentum tensor is dominated by 
a scalar field with specific potential. It is 
well known since the early seventies of the twentieth century that bouncing models 
may arise in the presence of spatial curvature without an explicit violation of the null energy condition \cite{boun5a}
but this is not the situation explored in this paper.}
$g_{\mu\nu} = a^{2}(\tau) \eta_{\mu\nu}$ where 
$\eta_{\mu\nu}= \mathrm{diag}(1, \, -1,\, -1, \, -1)$ is the Minkowski metric
 and $a(\tau)$ is the scale factor 
in the conformal time parametrization. In the case of a conformally flat metric the null vector $k^{\mu}$
can always be parametrized as $k^{\mu} = a^{-2} ( 1,\, n^{i})$ where $n_{i} n^{i} =1$. In this way 
$k^{\mu}$ coincides with the tangent to an affinely parametrized geodesic and Eq. (\ref{AC}) 
can be written as:
\begin{equation}
\int_{\gamma} T_{\mu\nu} k^{\mu} k^{\nu} \, d\lambda= \int_{\gamma} \,(p_{t} + \rho_{t}) \,d\tau \geq \, 0,
\label{three}
\end{equation}
where $(p_{t} + \rho_{t})$ denotes the total enthalpy density of the system and the equality follows by recalling 
that, in the present parametrization of null vectors $k^{\mu}$, $d\tau/d\lambda = a^{-2}$.
If the null energy condition is enforced for each and 
every value of the conformal time coordinate the bouncing dynamics is prevented.
We shall therefore be concerned with the situation where the null energy condition is violated in a limited 
range of the conformal time coordinate $\tau$ and, more specifically, between $-\tau_{-}$ and $\tau_{+}$. 
Indeed diverse solutions often employed to construct phenomenological 
scenarios can be divided in three distinct epochs: besides the intermediate time range (i.e. $- \tau_{-} <  \tau < \tau_{+}$) where the 
null energy condition is either partially or completely violated,  one can define 
two asymptotic regions (respectively for $\tau< - \tau_{-}$ and for $\tau>\tau_{+}$) where
the null, dominant and weak energy conditions are 
enforced\footnote{In spite of the conservative viewpoint of the present discussion, it could also be plausible 
to analyze, within the present framework, all those scenarios where the violation of the null energy condition 
occurs for $\tau\to -\infty$ (see e.g. the last paper of Ref. \cite{boun4}
for some examples along this direction); this potentially interesting generalizations, however, shall not be discussed here.}. 
With these specifications, the restrictions implied by Eq. (\ref{three}) can be rephrased as
\begin{equation} 
\biggl| \int_{- \tau_{-}}^{\tau_{+}} (p_{t} + \rho_{t}) \, d\tau \biggr| \leq  \int_{-\infty}^{- \tau_{-}} (p_{t} + \rho_{t}) \, d\tau + 
 \int_{\tau_{+}}^{+\infty} (p_{t} + \rho_{t})\, d\tau, 
\label{four}
\end{equation}
where, by definition, $(p_{t}+\rho_{t})_{\tau_{\pm}} \to 0$  while it is negative 
for $- \tau_{-} < \tau < \tau_{+}$.  Equation (\ref{four}) assumes that the curvature invariants are all regular and 
that the causal (i.e. the non-space-like) geodesics are all complete and extendible
 to arbitrary values of their affine parameter.

Since what matters is the total enthalpy density, Eqs. (\ref{three}) and (\ref{four}) do not solely depend on the dynamics of the 
background but also on the amplified inhomogeneities and, in particular, on those 
species which are not conformally coupled to the geometry. More specifically, if $N_{nc}$ 
is the number of non-conformally coupled species, it is well known that gravitational particle production at the end of inflation
roughly contributes to the total energy density as ${\mathcal N}_{nc} H_{f}^4(a_{f}/a)^4$ where $H_{f}$ is the 
Hubble rate at the end of inflation \cite{boun6} and ${\mathcal N}_{nc} = N_{nc} \,\zeta_{nc}$; the numerical factor $\zeta_{nc}< 1$ may change, 
in principle, from species to species. Following earlier and more recent estimates \cite{boun6} 
 we can have that $\zeta_{nc}= {\mathcal O}(0.01)$ for minimally coupled scalars so that, without loss of generality, we shall assume 
 ${\mathcal N}_{nc} = N_{nc}/100$. Similar values of $\zeta_{nc}$ arise when computing the total energy density 
 of gravitons since the two tensor polarizations (evolving in a conformally flat Friedmann-Robertson-Walker metric) obey 
the evolution of two independent scalar fields minimally coupled to the geometry \cite{boun6}. 
While the specific values of $\zeta_{nc}$ will be largely immaterial for the present purposes, 
the observations of Refs. \cite{boun6,boun7} (see also first paper of Ref. \cite{boun8}) 
cannot be disregarded whenever the rate of asymptotic expansion for $\tau \gg \tau_{+}$
is smaller than the expansion rate of radiation\footnote{It is interesting to remark that this observation has been used even 
prior to the formulation of the inflationary hypothesis: in a specific toy model
Parker demonstrated that the backreaction of massless quanta may lead to a thermal background radiation \cite{boun7}.}. 
In the latter case the energy density of the 
non-conformally coupled quanta may even become asymptotically dominant \cite{boun6}.
For the sake of concreteness we shall therefore assume that in the limits $\tau \to \mp \infty$ 
the background scale factor expands, respectively, as: 
 \begin{equation}
 \lim_{\tau \to - \infty} a(\tau) \to a_{-} \biggl(- \frac{\tau}{\tau_{-}}\biggr)^{\alpha_{-}/2}, \qquad 
 \lim_{\tau \to + \infty} a(\tau) \to a_{+} \biggl( \frac{\tau}{\tau_{+}}\biggr)^{\alpha_{+}/2}. 
 \label{five}
 \end{equation}
While the $\tau\to - \infty$ limit of the solution will only be determined by the evolution 
of the background, in the opposite limit the role of the produced massless quanta 
will depend on the value of $\alpha_{+}$; the total energy density for $\tau > \tau_{+}$ will be in fact given by: 
\begin{equation}
\rho_{t} = \frac{3 H_{+}^2 M_{P}^2}{8 \pi} \biggl(\frac{a_{+}}{a}\biggr)^{2 + 4 /\alpha_{+}} + {\mathcal N}_{nc} H_{+}^4 \biggl(\frac{a_{+}}{a}\biggr)^{4}.
\label{six}
\end{equation}
Note that ${\mathcal N}_{nc} \geq 1/50$ (where the equality holds if we just consider the two tensor polarizations).  
According to the parametrization of Eq. (\ref{six}), if $\alpha_{+} < 2$, the expansion rate will be
slower than in the case of radiation; conversely if $\alpha_{+} \geq 2$ the asymptotic rate of expansion will either 
be comparable or larger than the one typical of a radiation-dominated phase. This means that 
when $\alpha_{+} < 2$ the asymptotic contribution of the massless species will dominate; more specifically 
this will be the case for $a>a_{*}$ where $a_{+}/a_{*}$ is defined as:
\begin{equation}
\frac{a_{+}}{a_{*}} = 
\biggl(\frac{8 \pi {\mathcal N}_{nc}}{3}\biggr)^{\alpha_{+}/( 4 - 2 \alpha_{+})} \biggl(\frac{H_{+}}{M_{P}}\biggr)^{\alpha_{+}/( 2 -\alpha_{+})}.
\label{seven}
\end{equation}
In the opposite dynamical regime (i.e. $\alpha_{+} \geq 2$) the backreaction of the 
massless quanta is still conceptually relevant but does not have 
appreciable physical effects on the evolution of the background.

If $\alpha_{+} < 2$ the effect of the second term at the right hand side of  Eq. (\ref{six}) dominates the energy density for typical 
time-scales $\tau_{*} = {\mathcal O}(\tau_{+})$ whenever $H_{+} = {\mathcal O}(M_{P})$. 
Conversely the dominance of the amplified quantum fluctuations will take place on a time-scale much 
larger than the typical duration of the bouncing regime (i.e. $\tau_{*} \gg \tau_{+}$ 
whenever $H_{+}/M_{P} \ll 1$). For the sake of definiteness we can consider the standard Friedmann-Lema\^itre equations
\begin{equation}
3 M_{P}^2 \,{\mathcal H}^2 = 8 \pi \rho_{b} a^2, \qquad M_{P}^2 ({\mathcal H}^2  - {\mathcal H}^{\prime}) = 4 \pi (\rho_{b} + p_{b}) a^2,
\label{eight}
\end{equation} 
where $M_{P} = 1/\sqrt{G}$ denotes the Planck mass; a particular solution of Eq. (\ref{eight}) compatible with a local violation of the null energy condition 
and  expanding at a rate slower than radiation for $\tau> \tau_{+}$ is obtained in terms of the scale factor:
\begin{equation}
a(\tau) = a_{1} ( x^2 + 1)^{1/4}, \qquad x = \tau/\tau_{1},
\label{nine}
\end{equation}
leading, through Eq. (\ref{eight}), to the following energy and enthalpy densities\footnote{Note that 
the condition (\ref{AC}) is not bound to hold in flat space. It is however true that the examples examined in this paper 
for $\tau \to - \infty$, the background models go to flat space-time.}:
\begin{equation}
\rho_{b} = \frac{3 M_{P}^2}{32 \pi a_{1}^2 \tau_{1}^2} \frac{x^2}{(x^2 +1)^{5/2}}, \qquad \rho_{b} + p_{b} 
= \frac{ M_{P}^2}{16 \pi a_{1}^2 \tau_{1}^2} \frac{[3 x^2 - 2 ]}{(x^2 +1)^{5/2}}.
\label{ten}
\end{equation} 
Eq. (\ref{nine}) describes an accelerated contraction for $\tau  < -\tau_{-}$ turning into a decelerated expansion for $\tau > \tau_{+}$.
Equation (\ref{ten}) implies $\tau_{+} = \tau_{-} = \sqrt{2/3}\, \tau_{1}$ while $\alpha_{+} = \alpha_{-} = 1$. In the example 
of Eq. (\ref{nine}) the energy density of the non-conformally coupled species can be directly computed by 
studying the evolution equation of the corresponding mode functions. The production of massless 
quanta is governed (up to a numerical factor determining $\zeta_{nc}$) by $a^{\prime\prime}/a$  \cite{boun6,boun8}.
This function can be explicitly computed, for instance, in the case of Eq. (\ref{nine}) and it can be verified that 
all the Fourier modes $k < {\mathcal O}(1/\tau_{1})$ will indeed be amplified up to a  critical 
wavenumber\footnote{We recall that, as usual, ${\mathcal H}= a H$ where $H$ denotes the conventional Hubble rate.} $k_{max} = {\mathcal O}(a_{1} H_{1})$. The energy density of the produced quanta 
will then be dominated by the largest amplified mode and this is why it is 
of the order of $k_{max}^4 \simeq H_{1}^4 a_{1}^4$ \cite{boun6}.

Owing to the results of Eqs. (\ref{six}), (\ref{nine}) and (\ref{ten}) the explicit form of Eq. (\ref{three}) becomes:
\begin{equation}
\int_{-\infty}^{+\infty}\, (p_{t} + \rho_{t})  \, d\tau= \frac{M_{P} H_{1}^2}{16 \pi \tau_{1} a_{1}^2} 
\biggl[ \int_{- \infty}^{\tau_{*}/\tau_{1}} \frac{3 x^2 -2}{(x^2 +1)^{5/2}} \,  dx+
\frac{64 \pi}{9} {\mathcal N}_{nc} \biggl(\frac{H_{*}}{H_{1}}\biggr) \biggl(\frac{H_{*}}{M_{P}}\biggr)^2 
\biggl(\frac{a_{1}}{a_{*}}\biggr)\biggr].
\label{eleven}
\end{equation}
If we now recall that $H_{1} = {\mathcal H}_{1}/a_{1}$ we can also deduce, from Eqs. (\ref{six}) and (\ref{nine}), the following 
chain of equalities: 
\begin{equation}
\biggl(\frac{a_{1}}{a_{*}}\biggr)^2 = \frac{\tau_{1}}{\tau_{*}}= \frac{8 \pi}{3} {\mathcal N}_{nc} \biggl(\frac{H_{1}}{M_{P}}\biggr) = \epsilon,
\label{twelve}
\end{equation}
where, for the sake of conciseness, we introduced the dimensionless parameter $\epsilon$. 
Using  Eq. (\ref{twelve}), the result of Eq. (\ref{eleven}) can also be expressed as:  
\begin{equation}
\int_{-\infty}^{+\infty} \,  (p_{t} + \rho_{t})  \, d\tau= \frac{H_{1}^2}{ \tau_{1} a_{1}^2} \biggl[ \int_{- \infty}^{1/\epsilon}  \frac{3 x^2 -2}{(x^2 +1)^{5/2}}  dx+
\frac{8}{3} \epsilon^6\biggr];
\label{thirteen}
\end{equation}
to get rid of the $M_{P}^2/(16 \pi)$ in the prefactor of Eq. (\ref{thirteen}) we adopted units $16\pi G = 1$.
Equation (\ref{thirteen}) can be used within two complementary approaches which are, in short, the following. In a given bouncing 
scenario and for a specific value of ${\mathcal N}_{c}$ the value of $\epsilon$ can be accurately assessed: this 
analysis will show if (and how) the (averaged) null energy condition 
is enforced. Conversely, in a more heuristic perspective, we can use the positivity of the averaged null energy condition
to restrict the values of $\epsilon$. According to this second strategy,
the integral appearing in Eq. (\ref{thirteen}) can be performed analytically: the null energy condition is positive (in average) 
provided $0.8 < \epsilon < 1$  implying that $H_{1} \leq {\mathcal O}(M_{P})$. Thus by enforcing the 
averaged null energy condition  the maximal scale of the bounce can be determined.
Absent any contribution from the non-conformally coupled species, the integrals at the right hand side of  Eq. (\ref{four}) are subdominant 
if compared with the integral at the left hand side. However, when the backreaction effects are taken into account 
the second integral at the right hand side of Eq. (\ref{four}) becomes larger than the first contribution and 
this implies that the averaged null energy condition is not violated provided $H_{1} = {\mathcal O}(M_{P})$.

All in all we can say that, as long as $H_{1} \ll M_{P}$, the contribution of the non-conformally coupled species 
cannot restore the validity of the averaged null energy condition\footnote{ This statement follows from Eq. (\ref{thirteen}).
Unless $\epsilon$ is of order $1$ (but smaller than $1$) the expression at the right hand side of Eq. (\ref{thirteen})
is negative. }. The opposite is true in the case $H_{1} = {\mathcal O}(M_{P})$ 
that is the regime most efficiently described in terms stringy bounces \cite{boun8,boun9} rather than by  effective theories. 
For this purpose it is plausible to investigate how similar physical considerations arise 
in a class of bouncing models characterized by the contribution of a  nonlocal dilaton potential \cite{boun8}.
A related set of solutions can be realized in the context of double field theory \cite{boun9} which 
aims at realizing $T$-duality explicitly at the level of component fields of closed string field theory. 
In this class of bouncing universes \cite{boun8} the evolution equations in the Einstein frame are:
\begin{eqnarray}
&& 6 {\mathcal H}_{e}^2 = \frac{{ \varphi'}^2}{2}  + e^{\varphi} a_{e}^2 V,
\label{fourteenth}\\
&&  4 {\mathcal H}_{e}^{\prime} + 2 {\mathcal H}_{e}^2 = -\biggl(\frac{{\varphi}^{\prime\,2}}{2} - e^{\varphi} \,a_{e}^2 \, V \biggr) - e^{\varphi} a_{e}^2 \frac{\partial V}{\partial \overline{\varphi}},
\label{fifteenth}\\
&& \varphi^{\prime\prime} + 2 {\mathcal H}_{e} \varphi^{\prime} + e^{\varphi} a_{e}^2 \biggl( V - \frac{1}{2}  \frac{\partial V}{\partial \overline{\varphi}}\biggr)=0,
\label{sixteenth}
\end{eqnarray}
where, as in Eq. (\ref{thirteen}), natural gravitational units $16 \pi G=1$ have been adopted.
Equations (\ref{fourteenth})--(\ref{sixteenth}) have been derived in \cite{boun8} and can be found, 
exactly in this form, in the first paper of Ref. \cite{boun8}. The potential appearing in Eqs. (\ref{fourteenth})--(\ref{sixteenth}) 
depends on the $T$-duality invariant combination $\overline{\varphi}$, 
namely $V = V(\overline{\varphi})$ where\footnote{Note that $a_{s}$ is the scale factor in the string frame which is related to the scale factor 
in the Einstein frame as $a_{s} = a_{e} e^{\varphi/2}$; see first paper of Ref. \cite{boun8} for further details.}, 
in four space-time dimensions, $\overline{\varphi} = \varphi - 3 \ln{a_{s}}$. Since the potential depends on a $T$-duality invariant combination, 
the corresponding results can also be interpreted in a double field theory context (see, in particular, the last two papers of Ref. \cite{boun9}).

There are various classes of solutions compatible with a violation of the null energy condition in a finite time interval 
and some of them, with their own virtues and their potential drawbacks, have been discussed in the past (see e.g. 
Ref. \cite{boun8} and references therein). For the sake of concreteness we can consider the following potential: 
\begin{equation}
V(\overline{\varphi}) = V_{1} e^{2(\overline{\varphi} - \overline{\varphi}_{1})/\alpha} \biggl[ 1 - e^{2(\overline{\varphi} - 
\overline{\varphi}_{1})/\alpha}\biggr].
\label{seventeenth}
\end{equation}
After repeated combinations of Eqs.  (\ref{fourteenth}), (\ref{fifteenth}) and (\ref{sixteenth}) an explicit differential relation 
only involving ${\mathcal H}_{e}$ and $\varphi^{\prime}$ can be derived: 
 \begin{equation}
\frac{\partial}{\partial \tau} (\varphi^{\prime} + 2 {\mathcal H}_{e}) + 2 {\mathcal H}_{e} (\varphi^{\prime} + 2 {\mathcal H}_{e}) =0.
\label{eighteenth}
\end{equation}
Using now Eq. (\ref{eighteenth}) the bouncing solution corresponding to the potential (\ref{seventeenth}) is given by:
\begin{equation}
a_{e}(\tau) = a_{1} [ x^2 + 1]^{\alpha/4} , \qquad \varphi = \overline{\varphi}_{1} - \frac{\alpha}{2} \ln{(x^2 +1)}, \qquad x = \frac{\tau}{\tau_{1}},
\label{nineteenth}
\end{equation}
where, as in Eq. (\ref{nine}), $\tau_{1}$ denotes the typical scale of the bounce and ${\mathcal H} = a_{1} H_{1} = 1/\tau_{1}$. 
In the rescaled coordinate $x$ the bouncing region corresponds to $|x| <  \sqrt{2/(\alpha+2)}$ and to 
satisfy consistently all the equation the relation between the constants 
$\tau_{1}$, $V_{1}$ and $\overline{\varphi}_{1}$ must be given by $\tau_{1} a_{1} \sqrt{V_{1}} = \alpha e^{-\varphi_{1}/2}$ in natural 
gravitational units. 
The scale factor (\ref{nineteenth}) generalizes the one of Eq. (\ref{nine}) and 
the energy density and pressure, in the Einstein frame description, are:
\begin{equation}
\rho_{e} = \frac{{\varphi '}^2}{ 2 a_{e}^2} + e^{\varphi} V, \qquad p_{e} = \frac{{\varphi '}^2}{2 a_{e}^2} - e^{\varphi} V + 
e^{\varphi} \frac{\partial V}{\partial \overline{\varphi}}.
\label{twenty}
\end{equation}
In terms of $\rho_{e}$ and $p_{e}$ Eq. (\ref{sixteenth}) becomes $\rho_{e}' + 3 {\cal H}_{e} (\rho_{e} + p_{e}) =0$, as expected; finally the enthalpy 
density is
\begin{equation}
4( {\mathcal H}_{e}^2 - {\mathcal H}_{e}^{\prime}) = { \varphi'}^2 +   e^{\varphi} a_{e}^2 \frac{\partial V}{\partial \overline{\varphi}}.
 \label{twentyone}
 \end{equation}
Even if the evolution equations  (\ref{fourteenth})-- (\ref{sixteenth}) are local in time, the nonlocality of the potential which depends on the shifted dilaton 
$\overline{\varphi}$ is reflected in a substantially different form of the equations which cannot be mimicked\footnote{It is actually well known that if the potential 
only depends of $\varphi$ (e.g. $W= W(\varphi)$) we would simply have 
 $4({\mathcal H}_{e}^2 - {\mathcal H}_{e}^{\prime} )= \varphi^{\prime\,2}$ \cite{boun8}.}
 by a potential term depending only on $\varphi$.

As emphasized above, gravitational particle production at the end of inflation \cite{boun6,boun7} 
is caused by the presence of non-conformally coupled scalar fields.
While in the general theory of relativity the kinetic term of the gauge fields is conformally coupled to the background, 
in the case of stringy bounces gauge bosons can also be amplified thanks to the dilaton coupling; the quantum fluctuations of gauge 
fields will then contribute to the total energy density (see third paper of Ref. \cite{boun8}).  In this case, 
particle production  roughly contributes to the total energy density as ${\mathcal N}_{g} H_{1}^4(a_{1}/a)^4$ where, in analogy with Eq. (\ref{six}), ${\mathcal N}_{g}= N_{g} \zeta_{g}$; direct estimates of the evolution of the gauge mode functions suggest that $\zeta_{g} = {\mathcal O}(0.2)$ \cite{boun8} even if this value is largely irrelevant for the present considerations. This effect has been already 
 studied (both analytically and numerically) in a related context and for a slightly different background (see third paper of Ref. \cite{boun8}); in that case it has bee argued that particle production could heat up cold a bounce solution and, simultaneously, stabilize the dilaton. 
We are now ready to compute the averaged null energy condition in the case of the solution (\ref{nineteenth}).
More specifically the integral of the total enthalpy density becomes:
\begin{equation}
\int_{-\infty}^{+\infty} \, (p_{t} + \rho_{t}) \, d\tau = \frac{1}{4 \tau_{1} a_{1}^2} \biggl\{ \int_{-\infty}^{\tau_{*}} \frac{\alpha[(\alpha+2) x^2 -2]}{(x^2 +1)^{2 + \alpha/2}}\, d\tau
+\frac{16}{9} {\mathcal N}_{g} \tau_{*} a_{1} \frac{H_{*}^{4}}{H_{1}} \biggr\}.
\label{twentytwo}
\end{equation}
In the case of Eq. (\ref{twentytwo}) the analog of Eq. (\ref{twelve}) is given by:
\begin{equation}
\biggl(\frac{\tau_{1}}{\tau_{*}}\biggr) = \biggl[\frac{{\mathcal N}_{g}}{6} H_{1}^2\biggr]^{1/(2 -\alpha)} = \epsilon.
\label{twentythree}
\end{equation}
Using Eq. (\ref{twentythree}) we can rephrase Eq. (\ref{twentytwo}) as follows:
\begin{equation}
 \int_{-\infty}^{+\infty} \,  (p_{t} + \rho_{t})  \, d\tau= \frac{H_{1}}{a_{1}}\biggl\{ \int_{-\infty}^{\epsilon} \frac{\alpha[(\alpha+2) x^2 -2]}{(x^2 +1)^{2 + \alpha/2}}\, d x
 +\frac{8}{3} \epsilon^{\alpha + 5} \biggr\}.
 \label{twentyfour}
 \end{equation}
 The integral of Eq. (\ref{twentyfour}) can be performed explicitly in terms of hypergeometric functions so that the averaged null energy condition 
 evaluates to a function of $\alpha$ and $\epsilon$ which can be numerically studied. The averaged null energy condition is positive provided $0 < \alpha < 2$ and $0.5 < \epsilon < 1$. Consequently, as anticipated, the averaged null energy condition is not violated provided the maximal scale of the bounce is of the order 
 of (but smaller than) the Planck or string scale. 
 
If  all energy conditions are enforced throughout the dynamical evolution, non-singular
bouncing universes and their descendants must be largely excluded.
In this paper a less radical and somehow more conservative approach has been explored by 
suggesting a modest compromise: the null energy condition, even if locally violated, 
may be satisfied in an averaged sense. According to this prospect it has been 
assumed that the null energy condition is violated 
for a limited amount of time but not in the asymptotic regions (i.e. away from the bounce). 
Whenever the maximal expansion rate reached by the bounce 
is of the order of  Planck (or string) scales, the null energy condition is 
more likely to be satisfied in an averaged sense. Conversely, in a heuristic perspective, 
the enforcement of the (averaged) null energy condition furnishes an independent 
criterion for the determination of the maximal scale of the bounce. 

It is a pleasure to acknowledge A. Gentil-Beccot and S. Rohr of the CERN scientific information service 
for their kind assistance.

\end{document}